\documentstyle[12pt]{article}

\let\geqslant=\geq
\let\leqslant=\leq
\let\ov=\overline

\newcommand{\sect}[1]{\setcounter{equation}{0}\section{#1}}

\renewcommand{\thefootnote}{\fnsymbol{footnote}}

\def\appendix#1{
 \addtocounter{section}{1}
 \setcounter{equation}{0}
 \renewcommand{\thesection}{\Alph{section}}
 \section*{Appendix \thesection\protect\indent \parbox[t]{11.715cm} {#1}}
 \addcontentsline{toc}{section}{Appendix \thesection\ \ \ #1}
 }

\newcommand{\tr}[1]{\:{\rm tr}\,#1}

\def\scr{\scriptscriptstyle}
\def\dis{\displaystyle}
\def\e{{\rm e}\,}

\def\tw{\widetilde{W}}
\def\arctanh{\hbox{\,arctanh\,}}

\begin{document}
\begin{titlepage}
\begin{flushright}
SMI-Th-26/98\\
hep-th/9811200\\
\end{flushright}
\vspace{.5cm}

\begin{center}
{\LARGE  Two-logarithm matrix model with an external field}\\
\vspace{1.2cm}
{\large L. Chekhov\footnote{E-mail: chekhov@genesis.mi.ras.ru} }\\
\vspace{24pt}
{\it Steklov Mathematical Institute,}\\
{\it  Gubkina 8, 117966, GSP--1, Moscow, Russia}\\
\vspace{14pt}
and\\
\vspace{14pt}
{\large K. Palamarchuk\footnote{E-mail: palam@orc.ru} }\\
\vspace{24pt}
{\it Physical Department, Moscow State University,}\\
{\it Vorobyevy Gory, 119899, Moscow, Russia}\\
\end{center}
\vskip 0.9 cm
\begin{abstract}
We investigate the two-logarithm matrix model with the potential
$X\Lambda+\alpha\log(1+X)+\beta\log(1-X)$ related to an exactly
solvable Kazakov--Migdal model. In the proper
normalization, using Virasoro constraints, we prove
the equivalence of this model and the Kontsevich--Penner matrix model
and construct the $1/N$-expansion solution of this model.
\end{abstract}
\end{titlepage}
\setcounter{page}{1}
\renewcommand{\thefootnote}{\arabic{footnote}}
\setcounter{footnote}{0}

\sect{Introduction}

Matrix models with the coupling to external matrices plays an important role
in the contemporary mathematical and theoretical physics. Historically, the
first model of such type was the Brezin--Gross (BG)
model~\cite{BG} of the unitary
matrix~$U$ linearly coupled to an external matrix field~$\Lambda$. But the
real breakthrough in this field was caused by Kontsevich's papers~\cite{Konts}
where the generating functional for the 2D topological gravity was proved
to be the integral over the Hermitian matrices~$X$ with the potential $X^3$,
which are linearly coupled to an external matrix~$\Lambda$. Simultaneously,
the Witten hypothesis~\cite{Wit91} that this generating functional is a
$\tau$-function of the KdV hierarchy was proved~\cite{Konts,IZ92}.
The generalized Kontsevich model (GKM)---the model with an arbitrary
polynomial potential $V(X)$ and coupling with an external field---turned out
to be a $\tau$-function of the Kadomtsev--Petviashvili hierarchy~\cite{KMMM}.

Then, the interest to matrix models with logarithmic potentials appeared.
The first such model with the external field coupling was proposed
in~\cite{CM} (the authors named it the Kontsevich--Penner (KP) model)
and was pushed forward in~\cite{ACM,ACKM} where its
equivalence to the Hermitian one-matrix model with an arbitrary
nonsingular potential was proved. Underlying geometrical structure is
the discretized moduli space (d.m.s.) construction~\cite{Ch1}. Later on,
the exact relation was proved that connects this model in the d.m.s.\ times
with two copies of the Kontsevich integral taken at different time
sets~\cite{Ch2}.

Both the Kontsevich and the KP models, as well as the BG model, admit an
explicit solutions in the $1/N$-expansion~\cite{BG,IZ92,ACKM}. Such solutions
arise from the loop equation (or the Virasoro algebra constraints), which
are at most quadratic in fields. One can formulate the problem to find
{\it all} external field matrix models that manifest this property. Another
model of this kind was the so-called NBI matrix model of IIB superstrings
with the potential $X\Lambda+X^{-1}+(2\eta+1)\log X$ appeared~\cite{F-Z,ChZ}
in the context of the (M)atrix string theory. This model includes the BG
model as a particular case ($\eta=0$)~\cite{MMS} and away of this point,
it can be reduced~\cite{AC}, after the time changing, to the Kontsevich
model. (In particular, this enables one to produce the answer for the NBI
model in the moment technique as soon as the answer for the Kontsevich model
is known.) Note that the proof of equivalence of these two models relies
upon the coincidence of the Virasoro algebras.

The last model, which completes the list of matrix models with
the loop equations quadratic in fields and which can be therefore solved
in the $1/N$-expansion framework is the two-logarithm (2-log) model with the
potential $X\Lambda+\alpha\log(1-X)+\beta\log(1+X)$. This model turns out to
be closely related to the exactly solvable Kazakov--Migdal models~\cite{Mak1}
and it was thoroughly investigated in the case of the unit matrix~$\Lambda$,
i.e., where it is reduced to the one-matrix model. Even in this case, this
model manifests a rich phase structure~\cite{Mak2}.

In the present paper, we do not investigate all possible phases of the
2-log model and rather confine our consideration to the Kontsevich phase
only, in which the expansion over traces of negative powers of the
matrix~$\Lambda$ makes sense. First, we solve this model in the leading order
of the $1/N$-expansion; then, we find the constraint equations (the Virasoro
algebra) and prove that in the proper normalization, these equations are
exactly equivalent to the constraint equations of the KP model~\cite{CM}.
Possible applications of the 2-log model are discussed.

\sect{Matrix model with two logarithms}

We start with the following matrix integral, which appear, for instance,
in the logarithmic Kazakov--Migdal model~\cite{Mak1,Mak2}:
\begin{equation}\label{mm}
Z=\!\!\int\!\!dX\e^{-N\tr\left[X\Lambda
+\alpha\log (1-X) +\beta\log(1+X)\right]}\,.
\end{equation}
This integral is of the most general form, since, rescaling and
shifting the fields $X$ and $\Lambda$, we may change the logarithmic
branch points; however, we cannot change the constants $\alpha$ and
$\beta$, which are actual charges in the model\,\,(\ref{mm}).

The matrix integral\,\,(\ref{mm}) belongs to a class of generalized
Kontsevich models (GKM)~\cite{GKM}. Such models with negative powers of the
matrix~$X$ have been previously discussed in the context of
$c=1$ bosonic string theory~\cite{DMP}.

For the models of this type, the large $N$ solution is known explicitly only
in some special cases. The models with cubic potential for $X$~\cite{cubic}
and the combination of the logarithmic and quadratic potentials~\cite{CM,ZC}
were solved by a method based on the Schwinger--Dyson equations,
developed first for the unitary matrix models with external
field~\cite{BG,unit}. The same technique, being applied to the
integral\,\,(\ref{mm}), also allows one to find its large $N$ asymptotic
expansions in the closed form for arbitrary $\alpha$ and $\beta$.

The Schwinger--Dyson equations for\,\,(\ref{mm}) follow from the identity
\begin{equation}\label{...=0}
\left(\frac{1}{N^3}\,\frac{\partial }{\partial \Lambda_{jk} }
\,\frac{\partial }{\partial \Lambda_{li} }-\frac{1}{N}\delta_{jk}
\delta_{li}\right)\!\!\int\!\!dX
\frac{\partial }{\partial X_{ij}}\,
\e^{-N\tr\left[X\Lambda
+\alpha\log (1-X) +\beta\log(1+X)\right]}=0.
\end{equation}
Written in terms of the eigenvalues, these $N$ equations read
\begin{equation}\label{SDeig}
\left[-\frac{1}{N^2}\,\lambda_i\,\frac{\partial^2}{\partial
\lambda_i^2}-\frac{1}{N^2}\!\sum_{j\neq i}\lambda_j\,\frac{1}
{\lambda_j-\lambda_i}\left(\frac{\partial}{\partial\lambda_j}
-\frac{\partial}{\partial\lambda_i}\right)
+\frac{\alpha+\beta-2}{N}\,\frac{\partial}
{\partial\lambda_i}+(\beta-\alpha)+\lambda_i\right]\!Z(\lambda)\!=\!0.
\end{equation}

It is convenient to set
\begin{equation}\label{defW}
W(\lambda _i)=\frac{1}{N}\,\frac{\partial }{\partial \lambda _i}\,\log Z.
\end{equation}
We also introduce the eigenvalue density of the matrix $\Lambda $:
\begin{equation}\label{dens}
\rho (x)=\frac{1}{N}\,\sum_{i}\delta (x-\lambda _i).
\end{equation}
The density obeys the normalization condition
\begin{equation}\label{norm}
\int\!\!dx\,\rho(x)=1
\end{equation}
and in the large $N$ limit becomes a smooth function.

A simple power counting shows that the derivative of $W(\lambda_i )$ in
the first term on the left hand side of equation\,\,(\ref{SDeig})
is suppressed by the factor $1/N$ and can be omitted at $N=\infty $.
The remaining terms are rewritten as follows:
\begin{equation}\label{inteqn}
-\,xW^2(x)-\!\!\int\!\!dy\,\rho(y)\,y\,\frac{W(y)-W(x)}{y-x}
+(\alpha+\beta-2)W(x)+(\beta-\alpha)+x=0,
\end{equation}
where $\lambda_i$ is replaced by $x$. Equation\,\,(\ref{inteqn})
can be simplified by the substitution
\begin{equation}\label{subst}
\tw(x)=xW(x)-\frac{\alpha+\beta-1}{2}\,.
\end{equation}
After some transformations, using the normalization
condition\,\,(\ref{norm}), we obtain
\begin{equation}\label{maineq}
\tw^2(x)+x\!\!\int\!\!dy\,\rho(y)\,\,\frac{\tw(y)-\tw(x)}{y-x}=
x^2+(\beta-\alpha)x+\frac{(\alpha+\beta-1)^2}{4}\,.
\end{equation}

The nonlinear integral equation\,\,(\ref{maineq}) can be solved with the
help of the anzatz
\begin{equation}\label{anz}
\tw(x)=f(x)+\frac{x}{2}\!\int\!\!dy\,
\frac{\rho(y)}{f(y)}\,\frac{f(y)-f(x)}{y-x}\,,
\end{equation}
where $f(x)$ is an unknown function to be determined by
substituting\,\,(\ref{anz}) into Eq.\,(\ref{maineq}).
The asymptotic behaviors of $\tw(x)$ and
$f(x)$ as $x\rightarrow \infty $ follow from Eq.\,(\ref{maineq}):
$\tw(x)\sim {x}+(\beta-\alpha-1)/2$, and the analytic solution with
minimal set of singularities is merely
\begin{equation}\label{fx}
f(x)=\sqrt{ax^2+bx+c\,}.
\end{equation}
Let us introduce the moments of the external field
\begin{equation}
I_0=\!\!\int\!\!\frac{\rho(x)}{f(x)}\,dx,\qquad
J_0=\!\!\int\!\!\frac{\rho(x)}{f(x)}x\,dx.
\end{equation}
The parameters $a$, $b$, and $c$ are unambiguously determined
from Eq.\,(\ref{maineq}). We find that
$$
c=(\beta+\alpha-1)^2/4,
$$
and $a$ and $b$ are implicitly  defined by
the following two constraints:
\begin{eqnarray}\label{defa}
& &
1+\frac{1}{2}I_0=\frac{1}{\sqrt{a}}\,,
\nonumber
\\[-2.5mm]
& &
\\[-2.5mm]
& &
\sqrt{a}J_0=\beta-\alpha-\frac{b}{a}\,,
\nonumber
\end{eqnarray}
or, in terms of the eigenvalues,
\begin{eqnarray}\label{a1}
& &
1+\frac{1}{2N}\sum_{j}\frac{1}{\sqrt{a\lambda^2_j+
b\lambda_j+c\,}}=\frac{1}{\sqrt{a}}\,,
\nonumber
\\[-2.5mm]
& &
\\[-2.5mm]
& &
\sqrt{a}\frac{1}{N}\sum_{j}\frac{\lambda_j}{\sqrt{a\lambda^2_j+
b\lambda_j+c\,}}=\beta-\alpha-\frac{b}{a}\,.
\nonumber
\end{eqnarray}

So, we have
\begin{equation}\label{W1}
W(x)=\frac{\sqrt{ax^2+bx+c\,}}{x}+\frac{1}{2}\!\!\int\!\!dy\,\rho(y)\,
\frac{f(y)-f(x)}{f(y)(y-x)}+\frac{\alpha+\beta-1}{2x}\,.
\end{equation}
Then, integrating\,\,(\ref{W1})
w.r.t.\ $x$ and checking that the
stationary conditions w.r.t.\ the variables $a$ and $b$ take place,
we find the answer for the integral in the large $N$ limit,
\begin{eqnarray}\label{res}
& &
\hspace*{-7.0mm}\log Z=N^2(\beta-\alpha)^2\left[\frac{1}{8}
\log(b^2-4ac)-\frac{1}{4}\log a\right]+N^2(\beta-\alpha)
\Biggl[\frac{1}{4}\log a-\frac{1}{4}\log(b^2-4ac)
\nonumber
\\
& &
\hspace*{8.0mm}+\,\sqrt{c}\arctanh\frac{2\sqrt{ca}}{b}
-\frac{b}{2a}\Biggr]+N^2\left[\frac{b^2}{8a^2}-\frac{c}{2a}
+\frac{2c}{\sqrt{a}}+\frac{c}{2}\log(b^2-4ac)\right]
\nonumber
\\
& &
\hspace*{8.0mm}+\,N\sum_{i}\Biggl[\frac{\alpha+\beta-1}{2}\log\lambda_i
+\frac{f(\lambda_i)}{\sqrt{a}}
+\frac{1}{2}(\beta-\alpha)\log\left(\sqrt{a}\lambda_i
+\frac{b}{2\sqrt{a}}+f(\lambda_i)\right)
\nonumber
\\
& &
\hspace*{8.0mm}-\,\sqrt{c}\arctanh\biggl(\frac{\sqrt{c}
+\lambda_i\,b/(2\sqrt{c})}{f(\lambda_i)}\biggr)\Biggr]
-\frac{1}{4}\sum_{ij}\Biggl[\log(\lambda_i-\lambda_j)
\nonumber
\\
& &
\hspace*{8.0mm}+\,\arctanh\left(\frac{a\lambda_i\lambda_j
+(\lambda_i+\lambda_j)\,b/2+c}{f(\lambda_i)f(\lambda_j)}\right)\Biggr].
\end{eqnarray}
One can verify directly that
\begin{equation}\label{stat}
\frac{\partial}{\partial a}\log Z=
\frac{\partial}{\partial b}\log Z=0
\end{equation}
and $\dis{\frac{1}{N}\,\frac{\partial}{\partial\lambda_i}}\log
Z=W(\lambda _i)$, as far as Eq.\,(\ref{a1}) hold.

\sect{\large{\bf A large $N$ limit comparison}}

Let us establish a relation between the constraint equations
of the 2-log model and KP model~\cite{CM}.
It is convenient to introduce new charges (parameters) instead of
$\alpha$ and $\beta$
\begin{equation}
\gamma\equiv(\beta-\alpha)/2,\qquad
\varphi\equiv-\,(\alpha+\beta-1)/2=\sqrt{c},
\end{equation}
and new variables
\begin{equation}
\tilde b\equiv b/a,\qquad \tilde c\equiv c/a.
\end{equation}
Shifting all eigenvalues $\lambda_i$ by the same constant $\xi$,
we can rewrite the 2-log constraint equations as follows:
\begin{eqnarray}
& &
\frac{\varphi}{\sqrt{\tilde c}}\pm\frac{1}{2N}\sum_i
\frac{1}{\sqrt{\lambda_i^2+\left(2\xi+\tilde b\right)\lambda_i
+\left(\xi^2+\xi\tilde b+\tilde c\right)}}=1,
\nonumber
\\[-2.5mm]
& &
\label{2l_c_eq}
\\[-2.5mm]
& &
\pm\,\frac{1}{N}\sum_i\frac{\lambda_i+\xi}{\sqrt{\lambda_i^2
+\left(2\xi+\tilde b\right)\lambda_i+\left(\xi^2+\xi\tilde b
+\tilde c\right)}}=2\gamma-\tilde b,
\nonumber
\end{eqnarray}
where ``$\pm$" depends on the branch of the square root.
Then, we make the following time change:
\begin{equation}
{\rm tr}\frac{1}{\lambda^n}={\rm tr}\frac{1}{\eta^n}\pm
\biggl(-\,2\varphi\,\frac{N}{\left(-\xi\right)^n}+2N\delta_{n,1}
-N\delta_{n,2}\biggl),
\label{time1}
\end{equation}
where the role of ``$\pm$" is the same.
Making the presented time change
and connecting the variables of the two models
\begin{eqnarray}\label{var_2l_KP}
& &
2\xi+\tilde b=4b,
\nonumber
\\[-2.5mm]
& &
\\[-2.5mm]
& &
\xi^2+\xi\tilde b+\tilde c=4c,
\nonumber
\end{eqnarray}
where $b$ and $c$ are already the KP variables, we have
\begin{eqnarray}
& &
2b\pm\frac{1}{N}\sum_i\frac{1}{\sqrt{\eta_i^2+4b\eta_i+4c\,}}=0,
\nonumber
\\[-2.5mm]
& &
\label{KP_c_eq}
\\[-2.5mm]
& &
\pm\,\frac{1}{2N}\sum_i\frac{\eta_i}{\sqrt{\eta_i^2+4b\eta_i+4c\,}}
+c-3b^2=\gamma-\varphi,
\nonumber
\end{eqnarray}
i.e., exactly the constraint equations of the KP model with
$\gamma-\varphi\equiv\tilde\alpha$~\cite{CM}.
Here $\xi$ is an arbitrary parameter.
Using the original parameters $\alpha$, $\beta$, and
$\alpha_{\scr\rm KP}$ ($\alpha_{\scr\rm KP}$ is
the parameter $\alpha$ of the KP model,
and $\alpha_{\scr\rm KP}+1/2=\tilde\alpha$ in the notation of~\cite{CM}),
we see that $\alpha_{\scr\rm KP}=\beta-1$.

Naively, the parameter $\beta$ is more preferred than
$\alpha$ for some reason. Indeed, they play equal roles. The
obvious symmetry of the 2-log matrix integral is encoded in
the transformations $\lambda_i\rightarrow-\lambda_i$
($i=\overline{1,N}$) and $\alpha\leftrightarrow\beta$.
Under such a symmetry, $\gamma\rightarrow-\gamma$,
$\varphi\rightarrow\varphi$, and
$\alpha_{\scr\rm KP}=\gamma-\varphi\rightarrow\alpha_{\scr\rm KP}
=-\gamma-\varphi=\alpha-1$.

Let us recall the answer in the large $N$ limit for the KP model~\cite{CM}.
Substituting $\tilde\alpha$\,$=$\,$\gamma$\,$-$\,$\varphi$,
we have
\begin{eqnarray}
& &
\hspace*{-8.0mm}\log Z_{\scr\rm KP}
=\frac{N^2}{2}\left(\gamma-\varphi
-\frac{1}{2}\right)\log \left(b^2-c\right)-\frac{5}{2}\,b^2c
-\left(\gamma-\varphi\right) c
+\frac{c^2}{4}+3\left(\gamma-\varphi\right)b^2+\frac{9}{4}\,b^4
\nonumber
\\
& &
\hspace*{7.0mm}+\,N\sum_i\Biggl\{\left(\frac{\eta_i}{2}-b\right)
\sqrt{\frac{\eta_i^2}{4}+b\eta_i+c\,}
+\left(\gamma-\varphi\right)
\log \Biggl(\eta_i+2b+\sqrt{\frac{\eta_i^2}{4}+b\eta_i+c\,}\,\Biggr)
+\frac{\eta_i^2}{4}\Biggr\}
\nonumber
\\
& &
\hspace*{7.0mm}-\,\frac{1}{4}\sum_{ij}
\log \Biggl(\frac{\eta_i\eta_j}{4}+\frac{b}{2}\left(\eta_i+\eta_j\right)
+c+\sqrt{\frac{\eta_i^2}{4}+b\eta_i+c\,}\,
\sqrt{\frac{\eta_j^2}{4}+b\eta_j+c\,}\,\Biggr).
\label{g0_KP}
\end{eqnarray}
Here we compare the large $N$ limit answer for the 2-log
model with\,\,(\ref{g0_KP}). Further all equalities hold up to pure
complex constant and irrelevant factors which can
polynomially  depend only on the parameters $\alpha$ and $\beta$
(the polynomial of no more than second degree) of the 2-log model.
Obviously such additional terms cannot influence
the critical behavior of the model. Making the eigenvalue shift
by $\xi$ and denoting $d=b^2-c$, we obtain
\begin{eqnarray}
& &
\hspace*{-7.0mm}\log Z=\frac{N^2\gamma^2}{2}
\log d+N^2\gamma\Biggl[-\left(\varphi+\frac{1}{2}\right)
\log d+2\varphi\log\left(\tilde b
+2\sqrt{\tilde c}\right)-\tilde b\Biggr]
+N^2\Biggl[2d+2\varphi\sqrt{\tilde c}
\nonumber
\\
& &
\hspace*{5.0mm}+\,\frac{1}{2}\left(\varphi+\frac{1}{2}\right)^2
\log d-\varphi\left(\varphi+\frac{1}{2}\right)\log\tilde c\Biggr]
+N\sum_i\Biggl\{\gamma\log
\Biggl(1+\frac{2b}{\lambda_i}+\sqrt{1+\frac{4b}{\lambda_i}
+\frac{4c}{\lambda_i^2}\,}\,\Biggr)
\nonumber
\\
& &
\hspace*{5.0mm}-\,\varphi\log \Biggl(\sqrt{1+\frac{4b}{\lambda_i}
+\frac{4c}{\lambda_i^2}\,}+\frac{\tilde b}{2\sqrt{\tilde c\,}}
+\frac{\sqrt{\tilde c\,}
+\xi\,\tilde b/(2\sqrt{\tilde c\,})}{\lambda_i}\Biggr)
+\sqrt{\lambda_i^2+4b\lambda_i+4c\,}-\lambda_i
\nonumber
\\
& &
\hspace*{5.0mm}+\,\left(\gamma-\varphi-\frac{1}{2}\right)
\log \lambda_i+\lambda_i\Biggr\}
-\frac{1}{4}\sum_{ij}f\biggl(\frac{1}{\lambda_i}\,,
\frac{1}{\lambda_j}\biggr),
\end{eqnarray}
where
\begin{equation}
f\left(x,y\right)=\log \left(\frac{1}{2}+b\left(x+y\right)
+2cxy+\frac{1}{2}\sqrt{1+4bx+4cx^2\,}\,\sqrt{1+4by+4cy^2\,}\,\right).
\end{equation}

After some tedious algebra (similar to the one in~\cite{AC}), we obtain
\begin{equation}
\log Z=\log Z_{\scr\rm KP}+N\sum_i\Biggl\{\Bigl(\gamma
-\varphi-\frac{1}{2}\Bigr)
\log\,\Bigl(\frac{\lambda_i}{\eta_i}\Bigr)
+\lambda_i-\frac{\eta_i^2}{2}\Biggr\}
+2\varphi^2\log \varphi.
\label{NL_conn}
\end{equation}

The difference between $\log Z$ and $\log Z_{\scr\rm KP}$
depends only on some normalization factors in the large $N$ limit.
As is worth noting, these factors differ from the original
normalization factors of the two models, which can be obtained by
the early developed scheme~\cite{AC}.
We show that the appeared normalization factors are indeed natural.

Let us investigate the Kontsevich regime of the two models
($\Lambda\rightarrow\infty$ and $\eta\rightarrow\infty$).
Then, for KP model we have (up to a constant)
$$
Z_{\scr\rm KP}\!=\!\!\!\int\!\!DX \e
^{N{\rm tr}[\eta X\!-\!\frac{X^2}{2}+\alpha\log X]}=\e^{\frac{N}{2}
{\rm tr}\,\eta^2}({\rm det}\,\eta)^{\alpha N}\!\!\int\!\!DX
\e^{N{\rm tr}[-\frac{X^2}{2}+\alpha
\log(1+\frac{X}{\eta})]}\!\simeq\!\e^{\frac{N}{2}{\rm tr}\,\eta^2}
({\rm det}\,\eta)^{\alpha N}.
$$
There are two stationary points, $X_0=\pm 1+Y/\Lambda$, in the
Kontsevich regime for the 2-log model ($Y$ is the new variable).
Choosing $X_0=-1+Y/\Lambda$ for definiteness (another stationary
point gives the same answer after symmetry $\Lambda\rightarrow -\Lambda$),
we obtain
$$
Z=({\rm det}\,\Lambda)^{N(\beta-1)}\e^{N{\rm tr}\,\Lambda}
\!\!\int\!\! DY \e^{-N{\rm tr}[Y+\alpha\log(2-\frac{Y}{\Lambda})+\beta\log Y]}
\simeq({\rm det}\,\Lambda)^{N(\beta-1)}\e^{N{\rm tr}\,\Lambda}.
$$
This is nothing but our normalizing factors.

\sect{\large{\bf Constraint equations}}

Let us make the eigenvalue shift in the master equation of the 2-log model
$\Bigl(\partial_i\equiv\dis{\frac{\partial}{\partial\lambda_i}}\Bigr)$
\begin{equation}
\Bigl[-\,\frac{1}{N^2}(\lambda_i+\xi)\partial_i^2-\frac{1}{N^2}\sum_{j\neq i}
\frac{\lambda_j+\xi}{\lambda_j-\lambda_i}(\partial_j-\partial_i)
+\frac{\alpha+\beta-2}{N}\,\partial_i+\beta-\alpha+\lambda_i+\xi\Bigr]Z=0.
\end{equation}
Using our normalizing factor
\begin{equation}
\prod_i\lambda_i^{N(\beta-1)}e^{N\lambda_i}
\label{2l_n_f}
\end{equation}
and pushing it through derivatives, we replace
\begin{equation}
\partial_i\longrightarrow\partial_i+\frac{N(\beta-1)}{\lambda_i}+N.
\end{equation}
Then, we obtain master equation for the normalized partition function
\begin{eqnarray}
& &
\Bigl[-\,\frac{1}{N^2}(\lambda_i+\xi)\partial_i^2-\frac{1}{N^2}\sum_{j\neq i}
\frac{\lambda_j+\xi}{\lambda_j-\lambda_i}(\partial_j-\partial_i)
-\frac{2\lambda_i}{N}\,\partial_i+\frac{\alpha-\beta-2\xi}{N}\,\partial_i
\nonumber
\\
& &
\hspace*{2.0mm}-\,\frac{2\xi(\beta-1)}{N\lambda_i}\,\partial_i
-\frac{\xi(\beta-1)^2}{\lambda_i^2}+\frac{\beta-1}{\lambda_i}
\biggl(\alpha-2\xi+\frac{\xi}{N}\sum_j\frac{1}{\lambda_j}\biggr)
\Bigr]{\cal Z}=0.
\end{eqnarray}
Let us introduce the times of the 2-log model
\begin{equation}
t_n=\frac{1}{n}\sum_i\frac{1}{\lambda_i^n}\,.
\end{equation}
Then, the constraint equations for ${\cal Z}(\{t_n\})$ are obtained after
some tedious algebra which we omit here. Collecting all coefficients
to the term $1/(\lambda_i^kN^2)$, we obtain
\begin{equation}
L_k{\cal Z}(\{t_n\})=0,\qquad k\geqslant -1,
\end{equation}
where
\begin{eqnarray}
& &
L_k=V_{k+1}+\xi V_k+\xi N(\alpha+\beta-1)
\Bigl((1-\delta_{k,0}-\delta_{k,-1})\frac{\partial}{\partial t_k}
-N(\beta-1)\delta_{k,0}\Bigr)
\nonumber
\\
& &
\hspace*{10mm}+\,\xi\delta_{k,-1}N(\beta-1)(t_1-2N),
\end{eqnarray}
and
\begin{eqnarray}
& &
V_k=-\sum_{m=1}^{\infty}mt_m\frac{\partial}{\partial t_{m+k}}
-\sum_{m=1}^{k-1}\frac{\partial}{\partial t_m}\,
\frac{\partial}{\partial t_{k-m}}
-N(\alpha-\beta+1)(1-\delta_{k,0}-\delta_{k,-1})
\frac{\partial}{\partial t_k}
\nonumber
\\
& &
\hspace*{10.0mm}+\,2N(1-\delta_{k,-1})\frac{\partial}{\partial t_{k+1}}
+t_1\delta_{k,-1}\frac{\partial}{\partial t_{k+1}}
+N^2\alpha(\beta-1)\delta_{k,0}\,.
\end{eqnarray}
Here, the derivatives over $t_0$ and $t_{-1}$ are fictitious
and are used only for compactifying the presentation.

For $k,l\geqslant -1$, $L_k$ satisfy the algebra
\begin{equation}
[L_k,L_l]=(l-k)(L_{k+l+1}+\xi L_{k+l}).
\end{equation}
Zero shift ($\xi=0$) results in the Virasoro algebra
where the $L_{-1}$ generator is absent,
\begin{equation}
[V_k,V_l]=(l-k)V_{k+l}, \qquad k,l\geqslant 0.
\end{equation}
 We can also obtain the Virasoro algebra from the general algebra
with nonzero shift by the replacement
\begin{equation}
{\cal L}_k=\sum_{s=0}^{\infty}\frac{(-1)^s}{\xi^{s+1}}L_{k+s},
\qquad  k\geqslant -1,
\end{equation}
which is singular at $\xi=0$.
Performing the replacement and using the relations
$\alpha_{\scr\rm KP}=\beta-1$ and $\varphi=-(\alpha+\beta-1)/2$,
we obtain
\begin{eqnarray}
& &
{\cal L}_k=-\sum_{m=1}^{\infty}mt_m\frac{\partial}{\partial t_{m+k}}
-\sum_{m=1}^{k-1}\frac{\partial}{\partial t_m}\,
\frac{\partial}{\partial t_{k-m}}
+2N\alpha_{\scr\rm KP}\frac{\partial}{\partial t_k}
+2N\frac{\partial}{\partial t_{k+1}}
\nonumber
\\
& &
\hspace*{10.0mm}-\,2\varphi N\sum_{s=1}^{\infty}\frac{1}{(-\xi)^s}\,
\frac{\partial}{\partial t_{k+s}}
-2N\alpha_{\scr\rm KP}(\delta_{k,0}+\delta_{k,-1})
\frac{\partial}{\partial t_k}
-N^2\alpha_{\scr\rm KP}^2\delta_{k,0}
\nonumber
\\
& &
\hspace*{10.0mm}+\,\delta_{k,-1}
\biggl(t_1-2N-\frac{2\varphi N}{\xi}\biggr)\biggl(N\alpha_{\scr\rm KP}+
\frac{\partial}{\partial t_{k+1}}\biggr).
\end{eqnarray}
After the time changing
\begin{equation}
t_n=\tilde t_n -2\varphi\frac{N}{(-\xi)^n}+2N\delta_{n,1}
-\frac{N}{2}\delta_{n,2}\,,
\label{time2}
\end{equation}
where
\begin{equation}
\tilde t_n=\frac{1}{n}\sum_i\frac{1}{\eta_i^n}
\end{equation}
are the times of the KP model, we obtain
\begin{eqnarray}
& &
\hspace*{-10.0mm}{\cal L}_k=-\sum_{m=1}^{\infty}m\tilde t_m
\frac{\partial}{\partial\tilde t_{m+k}}
-\sum_{m=1}^{k-1}\frac{\partial}{\partial\tilde t_m}\,
\frac{\partial}{\partial\tilde t_{k-m}}
+2N\alpha_{\scr\rm KP}\frac{\partial}{\partial\tilde t_k}
+N\frac{\partial}{\partial\tilde t_{k+2}}
\nonumber
\\
& &
-\,2N\alpha_{\scr\rm KP}(\delta_{k,0}+\delta_{k,-1})
\frac{\partial}{\partial\tilde t_k}
-N^2\alpha_{\scr\rm KP}^2\delta_{k,0}
+\tilde t_1\delta_{k,-1}\frac{\partial}{\partial\tilde t_{k+1}}
+N\alpha_{\scr\rm KP}\tilde t_1\delta_{k,-1}\,.
\end{eqnarray}

This is exactly the Virasoro algebra that appears in the KP model with
the normalizing factor
\begin{equation}
\prod_i \eta_i^{\alpha_{\scr\rm KP}N}
e^{\frac{N}{2}\eta_i^2}\,.
\label{KP_n_f}
\end{equation}
Indeed, we can perform the same operation for the KP model. First, we write
the master equation for the normalized partition function
$\Bigl(\partial_i\equiv\dis{\frac{\partial}{\partial\eta_i}}\Bigr)$
\begin{equation}
\Bigl[-\,\frac{1}{N^2}\partial_i^2-\frac{1}{N^2}\sum_{j\neq i}
\frac{\partial_j-\partial_i}{\eta_j-\eta_i}-\frac{\eta_i}{N}\,\partial_i
-\frac{2\alpha_{\scr\rm KP}}{N\eta_i}\,\partial_i+
\frac{\alpha_{\scr\rm KP}}{N\eta_i}\sum_j\frac{1}{\eta_j}
-\frac{\alpha_{\scr\rm KP}^2}{\eta_i^2}\Bigr]{\cal Z}_{\scr\rm KP}=0.
\end{equation}
Then, using the KP model times $\tilde t_n$ and collecting all coefficients
to the term $1/(\eta_i^k N^2)$, we obtain
\begin{equation}
{\cal L}_k{\cal Z}_{\scr\rm KP}=0,\qquad k\geqslant -1.
\end{equation}
Therefore, we have proven the equivalence between the 2-log and KP models.

Now, we write the explicit relation between the normalized partition
functions of the two models
\begin{equation}
{\cal Z}_{\scr\rm 2\mbox{-}log}\Bigl[\biggl\{\frac{1}{n}\,
{\rm tr}\frac{1}{\lambda^n}\biggr\};\alpha,\beta\Bigr]
=C(\alpha,\beta)\xi^{2\varphi(\beta-1)N^2}\e^{N^2(2\beta-1)\xi}
{\cal Z}_{\scr\rm KP}\left[\tilde t_n(\xi,\varphi),
\alpha_{\scr\rm KP}\right],
\label{ex_rel}
\end{equation}
where
\begin{eqnarray}
& &
{\cal Z}_{\scr\rm 2\mbox{-}log}\Bigl[\biggl\{\frac{1}{n}\,
{\rm tr}\frac{1}{\lambda^n}\biggr\};\alpha,\beta\Bigr]
=\frac{Z_{\scr\rm 2\mbox{-}log}\left[\lambda;\alpha,\beta\right]}
{\prod_i\left\{(\lambda_i-\xi)^{N(\beta-1)}\,
\e^{N(\lambda_i-\xi)}\right\}}\,,
\nonumber
\\[-2.5mm]
& &
\\[-2.5mm]
& &
{\cal Z}_{\scr\rm KP}\left[\tilde t_n(\xi,\varphi),
\alpha_{\scr\rm KP}\right]
=\frac{Z_{\scr\rm KP}\left[\eta(\xi,\varphi),
\alpha_{\scr\rm KP}\right]}
{\prod_i\left\{(\eta_i)^{N\alpha_{\scr\rm KP}}\,
\e^{\frac{N}{2}\eta_i^2}\right\}}\,,
\nonumber
\end{eqnarray}
$\alpha_{\scr\rm KP}$\,$=$\,$\beta-1$ and $C(\alpha,\beta)$ is
some constant depending on the parameters $\alpha$ and $\beta$.

Note that we use here unshifted initial field $\lambda$ and
explicitly show the dependence on the arbitrary parameter $\xi$ by
the following reason. For the unshifted $\lambda$-field,
the Virasoro algebra for the 2-log model does not possess
the $L_{-1}$ generator. So, a question arises how we can obtain
the ${\cal L}_{-1}$ generator of the KP model when passing to
the KP model. The reason is that after the time
changing\,\,(\ref{time2}), the KP times $\tilde t_n$
become $\xi$-dependent.
Differentiating\,\,(\ref{ex_rel}) over $\xi$
and using the relation
\begin{equation}
\frac{d\tilde t_n}{d\xi}=(n+1)\tilde t_{n+1}-N\delta_{n,1}\,,
\end{equation}
we obtain one more equation for ${\cal Z}_{\scr\rm KP}$\,,
$$
{\cal L}_{-1}{\cal Z}_{\scr\rm KP}=0,
$$
where ${\cal L}_{-1}$ is just the generator of the KP Virasoro algebra.

\sect{\large{\bf Higher genus expressions}}

Let us recall the genus expansion for the KP model~\cite{ACKM},
\begin{equation}
\log {\cal Z}_{\scr\rm KP}=\sum_{g=0}^\infty N^{2-2g}F_g\,,
\end{equation}
where
\begin{equation}
F_g=\!\!\!\sum_{\alpha_j>1,\,\beta_j>1}\!\!\!
\left\langle\alpha_1\dots\alpha_s;
\beta_1\dots\beta_l|\alpha\beta\gamma\right\rangle_g
\frac{M_{\alpha_1}\dots M_{\alpha_s}J_{\beta_1}\dots J_{\beta_l}}
{M_1^\alpha J_1^\beta d^\gamma}\,,\qquad g>1,
\end{equation}
and
\begin{equation}
F_1=-\,\frac{1}{24}\log(M_1J_1d^4),\qquad g=1.
\end{equation}
The moments were defined as follows ($k\geqslant 0$)
\begin{eqnarray}
& &
M_k=\frac{1}{N}\sum_{i=1}^{N}\frac{1}{(\eta_i-x_+)^{k+1/2}\,
(\eta_i-x_-)^{1/2}}-\delta_{k,1}\,,
\nonumber
\\[-2.5mm]
& &
\\[-2.5mm]
& &
J_k=\frac{1}{N}\sum_{i=1}^{N}\frac{1}{(\eta_i-x_+)^{1/2}\,
(\eta_i-x_-)^{k+1/2}}-\delta_{k,1}\,,
\nonumber
\end{eqnarray}
where $x_{\pm}$ are the endpoints of the cut for the one-cut solution and
$d=x_+-x_-$\,. In our notation,
\begin{equation}
x_{\pm}=-2b\pm\sqrt{4b^2-2c\,}\,.
\end{equation}

Let us introduce the moments for the 2-log model ($k\geqslant 0$)
\begin{eqnarray}
& &
N_k=\frac{1}{N}\sum_{i=1}^{N}\frac{1}{(\lambda_i-y_+)^{k+1/2}\,
(\lambda_i-y_-)^{1/2}}\,,
\nonumber
\\[-2.5mm]
& &
\\[-2.5mm]
& &
K_k=\frac{1}{N}\sum_{i=1}^{N}\frac{1}{(\lambda_i-y_+)^{1/2}\,
(\lambda_i-y_-)^{k+1/2}}\,,
\nonumber
\end{eqnarray}
where
\begin{equation}
y_{\pm}=-\,\frac{\tilde b}{2}\pm\sqrt{\frac{\tilde b^2}{4}-\tilde c\,}\,.
\end{equation}
We are interested in the relation between the moments of the two models for
$k$\,$\geqslant$\,$0$ (for $k$\,$=$\,$0$, the relation is given
by constraint equations\,\,(\ref{2l_c_eq}) and\,\,(\ref{KP_c_eq})).
After making the eigenvalue shift ($y_{\pm}$\,$=$\,$x_{\pm}+\xi$)
and performing the time changing,
we obtain ($k\geqslant 1$)
\begin{eqnarray}
& &
M_k=N_k+2\varphi\frac{(-1)^{k+1}}{y_+^{k+1/2}\,y_-^{1/2}}\,,
\nonumber
\\[-2.5mm]
& &
\\[-2.5mm]
& &
J_k=K_k+2\varphi\frac{(-1)^{k+1}}{y_+^{1/2}\,y_-^{k+1/2}}\,.
\nonumber
\end{eqnarray}

So, for the 2-log model, we have
\begin{equation}
\log {\cal Z}=\sum_{g=0}^\infty N^{2-2g}F_g^{\scr\rm 2\mbox{-}log}\,,
\end{equation}
where
\begin{eqnarray}
& &
F_g^{\scr\rm 2\mbox{-}log}=\!\!\!\sum_{\alpha_j>1,\,\beta_j>1}\!\!\!
\left\langle\alpha_1\dots\alpha_s;
\beta_1\dots\beta_l|\alpha\beta\gamma\right\rangle_g
\,\prod_{i=1}^{s}\biggl(N_{\alpha_i}+2\varphi
\frac{(-1)^{\alpha_i+1}}{y_+^{\alpha_i+1/2}\,y_-^{1/2}}\biggr)
\nonumber
\\
& &
\hspace*{20.0mm}\times\prod_{i=1}^{l}\biggl(K_{\beta_i}+2\varphi
\frac{(-1)^{\beta_i+1}}{y_+^{1/2}\,y_-^{\beta_i+1/2}}\biggr)
\Bigl\{\biggl(N_1+2\varphi\frac{1}{y_+^{3/2}\,y_-^{1/2}}\biggr)^\alpha
\nonumber
\\
& &
\hspace*{20.0mm}\times\biggl(K_1+2\varphi
\frac{1}{y_+^{1/2}\,y_-^{3/2}}\biggr)^\beta
(y_+-y_-)^\gamma\Bigr\}^{-1},\qquad g>1,
\label{2l_exp_g}
\end{eqnarray}
and
\begin{equation}
F_1=-\,\frac{1}{24}\log
\Bigl\{\biggl(N_1+2\varphi\frac{1}{y_+^{3/2}\,y_-^{1/2}}\biggr)
\biggl(K_1+2\varphi\frac{1}{y_+^{1/2}\,y_-^{3/2}}\biggr)
(y_+-y_-)^4\Bigr\}.
\label{2l_exp_1}
\end{equation}

Therefore, expression\,\,(\ref{res}) for genus zero, taking into
account the normalizing factor\,\,(\ref{2l_n_f}), and
expressions\,\,(\ref{2l_exp_g}), (\ref{2l_exp_1}),
completely determine the partition function of the model\,\,(\ref{mm})
at all genera.

\sect{\large{\bf Determinant formulas}}

The exact determinant formulas in our model
can be easily found using the Itzykson--Zuber--Mehta technique
for the integration over angular variables in multi-matrix models.
The partition function can be expressed as follows
\begin{equation}
Z=(2\pi)^{\frac{N^2-N}{2}}\!\int\limits_{\theta_1}^{\theta_2}\!
\prod_i\left\{dx_i(1-x_i)^{-\alpha N}
(1+x_i)^{-\beta N}\e^{-N\lambda_i x_i}\right\}
\frac{\triangle(x)}{\triangle(\lambda)}\,,
\end{equation}
where $\triangle(x)=\prod_{i>j}^{N}(x_i-x_j)$
is the Van der Monde determinant and $\theta_{1,2}$ are some
integration limits. We know that in the large $N$ limit,
the difference between partition functions calculated
in various integration limits is exponentially small and does not affect
the $1/N$ perturbative expansion. So, we investigate several cases.

(i).\,\,\,For $\theta_1=-1$ and $\theta_2=1$, we use
the following integral representation ($a,b>0$)
\begin{equation}
\int\limits_{-1}^1\!\!dx\,(1-x)^{a-1}(1+x)^{b-1}\e^{-cx}
=2^{a+b-1}\,\e^{-c}\,B(a,b)\,\Phi(a,a+b;2c),
\end{equation}
where $\Phi(a,c;z)$\,$\equiv$\,$_1F_1(a,c;z)$
is the confluent hypergeometric
function and $B(a,b)$ is the beta-function.

Then, in the domain $\alpha,\beta<1/N$, we obtain
\begin{eqnarray}
& &
Z=(2\pi)^{\frac{N^2-N}{2}}2^{-(\alpha+\beta)N^2+N(N+1)/2}
\prod_i \biggl\{B(-\alpha N+1,-\beta N+i)\biggr\}
\nonumber
\\
& &
\hspace*{9.0mm}\times\,\frac{\e^{-N{\rm tr}\,\lambda}}{\triangle(\lambda)}
\det_{1\leqslant i,\,j\leqslant N}||
\Phi(-\alpha N+1,-(\alpha+\beta)N+j+1;2N\lambda_i)||\,.
\label{det_1}
\end{eqnarray}

(ii).\,\,\,For $\theta_1=1$ and $\theta_2=\infty$, we use
the relation ($a,c>0$)
\begin{equation}
\int\limits_1^{\infty}\!\!dx\,(1-x)^{a-1}(1+x)^{b-1}\e^{-cx}
=(-1)^{a-1}2^{a+b-1}\,\e^{-c}\,\Gamma(a)\,\Psi(a,a+b;2c),
\end{equation}
where
\begin{equation}
\Psi(a,c;z)=\frac{\Gamma(1-c)}{\Gamma(a-c+1)}\,\Phi(a,c;z)
+\frac{\Gamma(c-1)}{\Gamma(a)}\,z^{1-c}\,\Phi(a-c+1,2-c;z)
\end{equation}
is the other confluent hypergeometric function and
$\Gamma(a)$ is the gamma-function.

Then, in the domain where $\alpha<1/N$, $\beta$
is unrestricted, and $\lambda_i>0$, we have
\begin{eqnarray}
& &
Z=(2\pi)^{\frac{N^2-N}{2}}(-1)^{-\alpha N^2}
2^{-(\alpha+\beta)N^2+N(N+1)/2}\,\Gamma^N(-\alpha N+1)
\nonumber
\\
& &
\hspace*{9.0mm}\times\,\frac{\e^{-N{\rm tr}\,\lambda}}{\triangle(\lambda)}
\det_{1\leqslant i,\,j\leqslant N}||
\Psi(-\alpha N+1,-(\alpha+\beta)N+j+1;2N\lambda_i)||\,.
\end{eqnarray}
This answer covers more general domain of the parameters
$\alpha$ and $\beta$ than\,\,(\ref{det_1}).

(iii).\,\,\,If $\alpha=0$ we get the simplest
answer setting $\theta_1=-1$ and $\theta_2=\infty$.
In the domain $\beta<1/N$ and $\lambda_i>0$, we obtain
\begin{equation}
\Bigl.Z\Bigr|_{\alpha=0}=
(2\pi)^{\frac{N^2-N}{2}}\prod_i\left\{
\Gamma(-\beta N+i)\right\}
(\det\Lambda)^{(\beta-1)N}\e^{N{\rm tr}\,\Lambda}\,,
\end{equation}
which is the unshifted normalizing factor up to a constant.

\sect{\large{\bf String susceptibilities}}

Let us calculate the string susceptibility w.r.t.\ $\gamma$ and $\varphi$
for\,\,(\ref{res}). By virtue of Eq.\,(\ref{stat}),
$\dis{\frac{d}{d\gamma}}\log Z=\dis{\frac{\partial}{\partial\gamma}}\log Z$
and the same holds true for $\varphi$.
Furthermore, an amazing fact is that the expressions obtained
are themselves stationary w.r.t.\ differentiation over $a$ and $b$.
This means that the total second derivatives in $\gamma$
and $\varphi$ coincide with the corresponding partial derivatives, so we have
\begin{eqnarray}\label{suscept}
& &
\chi_1=\frac{1}{N^2}\frac{d^2}{d\gamma^2}\log Z=\log(b^2-4ac)-2\log a\,,
\nonumber
\\
& &
\chi_2=\frac{1}{N^2}\frac{d^2}{d\gamma d\varphi}\log Z=
-2\arctanh\frac{2\sqrt{ac}}{b}=
-\log\frac{b+2\sqrt{ac}}{b-2\sqrt{ac}}\,,
\\
& &
\chi_3=\frac{1}{N^2}\frac{d^2}{d\varphi^2}\log Z=\log(b^2-4ac)+6.
\nonumber
\end{eqnarray}

Recalling the string susceptibility of the KP model
in the KP variables $b$ and $c$~\cite{CM}
\begin{equation}
\chi_{\scr\rm KP}=\log(b^2-c)
\end{equation}
and using the relations\,\,(\ref{var_2l_KP}), we obtain
\begin{equation}
\chi_{\scr\rm KP}=\chi_1.
\end{equation}

\sect{Conclusion}
This paper concludes the series of papers~\cite{CM,Ch2,ChZ,AC}
devoted to styduing the external field matrix problems with
logarithmic potentials. We see that, at least in the $1/N$-expansion
in terms of the corresponding moments,
all these models can be reduced either to the Kontsevich model or to
the Hermitian one-matrix model with an arbitrary potential. Here,
the question arises whether this can be derived directly within the
$\tau$-function framework~\cite{KMMM}. The related question is
which reductions of the Kadomtsev--Petviashvili hierarchy correspond
to the NBI and 2-log models.

One can always say the origin of the logarithmic terms is due to
additional degrees of freedom that were integrated out. Matrix integral
(\ref{mm}) can be represented as the $O(\alpha,\beta)$-type~\cite{O(n)}
matrix integral
\begin{equation}
\label{mmOn}
Z=\!\!\int\!\!dX\,\prod_{i=1}^{\alpha}d{\ov \Psi}_i d\Psi_i
\prod_{j=1}^{\beta}d{\ov \Phi}_j d\Phi_j
\e^{-N\tr\left[{\ov \Psi}_i \Psi_i +{\ov \Phi}_j \Phi_j
+X\bigl(\Lambda-\Psi_i {\ov \Psi}_i+\Phi_j {\ov \Phi}_j\bigr)\right]}\,,
\end{equation}
where the sum over repeated indices is implied and we assume the
matrix fields~$\Phi$ and~$\Psi$ are Grassmann even. Action (\ref{mmOn})
is of a nonlinear sigma model type with free matrix fields~$\Phi$ and~$\Psi$
dwelling on the manifold $\Lambda-\Psi_i {\ov \Psi}_i+\Phi_j {\ov \Phi}_j=0$.

\sect{Acknowledgements}
The work is supported by the Russian Foundation for Basic Research
(Grant No.~96--01--00344).

\end{document}